\documentclass[aps,pre,reprint,groupedaddress,noofootinbib]{revtex4-1}

\usepackage{chemarrow}
\usepackage{url}
\usepackage{graphicx}
\usepackage{gensymb}

\usepackage{ulem} 
\usepackage{cancel} 

\begin{document}

\normalem 

\title{Computing equilibrium concentrations for large hetero-dimerization networks}

\author{M. G. A. {van Dorp}}
\affiliation{Institute for Theoretical Physics, KULeuven,
Celestijnenlaan 200D, B-3001 Leuven, Belgium}

\author{F. Berger}
\affiliation{Institute for Theoretical Physics, KULeuven,
Celestijnenlaan 200D, B-3001 Leuven, Belgium}

\author{E. Carlon}
\affiliation{Institute for Theoretical Physics, KULeuven,
Celestijnenlaan 200D, B-3001 Leuven, Belgium}

\begin{abstract}
We consider a chemical reaction network governed by mass action kinetics
and composed of $N$ different species which can reversibly form heterodimers.
A fast iterative algorithm is introduced to compute the equilibrium
concentrations of such networks. We show that the convergence is
guaranteed by the Banach fixed point theorem. As a practical example, of
relevance for a quantitative analysis of microarray data, we consider
a reaction network  formed by $N \sim 10^6$ mutually hybridizing
different mRNA sequences. We show that, despite the large number of
species involved, the convergence to equilibrium is very rapid for
most species.
The origin of slow convergence for some specific subnetworks is discussed. 
This provides some insights for improving the performance of the algorithm.
\end{abstract}

\pacs{82.20.-w,05.10-a,82.39-k}
\date{\today}

\maketitle

\section{Introduction}

Systems of coupled chemical reactions involving many different
species, i.e. reaction networks, have been intensively studied in
chemistry, physics, mathematics, and engineering sciences (see e.g.
\cite{fein87,dejo02,gerd04,conr05,alon06,mart10,shin10,zwic10,pant10,mill11}).
If diffusion is fast enough and the number of molecules is sufficiently
large, so that stochastic effects can be neglected, these systems can
be described by a set of coupled first order ordinary differential
equations (ODE), which govern the time evolution of the concentrations
of each species.  In the ODE description, the rates of production and
consumption of the chemical species are given in term of mass action,
Michaelis-Menten, or other cooperative-type kinetics \cite{dejo02}.
In such systems different types of behavior are possible, as for instance
relaxation to a unique stationary point, oscillations or multistability.

Usually the time evolution of the system can be computed through numerical
integration of the ODE. However, this method can become very slow for
large reaction networks. In addition, the main interest is typically the
long time behavior of the system, which in absence of oscillations boils
down to finding the stationary (equilibrium) concentrations of each of
the chemical species.

In the present work, we will describe an efficient method to find
equilibrium concentrations for a class of networks which we will refer to as
hetero-dimerization networks. In these networks the species associate to form
dimers, which eventually break apart giving back the single species. The
method is based on an iterative scheme, of which we can rigorously
prove the convergence. The proof relies on the Banach fixed point theorem.

The proposed method is very efficient for large reaction networks. As an
example to show that convergence is fast, even for systems  with $\sim
10^6$ species, we consider the hybridization of RNA strands. This
example is of relevance, for instance, for a better quantitative
understanding of the reactions underlying the functioning of DNA
microarrays \cite{carl06,bind06,burd06,halp06}. It shows that some
sequences tend to get effectively depleted from the solution because
of partial complementarity with other strands. This brings some
consequences for the design and interpretation of microarray experiments.

This paper is organized as follows. In Section~\ref{sec:dimer} we
introduce the iterative algorithm and prove its convergence to the fixed
point, irrespective of the initial condition. In Section~\ref{sec:mRNA}
we show a concrete calculation for a network composed of a large number
of hybridizing mRNA strands.  Section~\ref{sec:convergence} discusses
the convergence rate of the algorithm and Section~\ref{sec:conclusion}
concludes the paper.

\section{Hetero-dimerization reactions}
\label{sec:dimer}

\subsection{Iterative scheme for stationary point}

We consider a set of different chemical species $A_i$ ($i=1,2 \ldots N$)
undergoing reversible association/dissociation reactions of the type:
\begin{equation}
A_i + A_j \autorightleftharpoons{$K_{ij}$}{$\overline{K}_{ij}$} A_{ij}.
\label{reactions}
\end{equation}
where $K_{ij}$ and $\overline{K}_{ij}$ are the forward and reverse rates.
The ratio of the rates must satisfy the detailed balance condition
\begin{equation}
K_{ij}/\overline{K}_{ij} = e^{\Delta G_{ij}/RT}
\label{DB}
\end{equation}
where $\Delta G_{ij}$ is the free energy of formation of the complex $A_{ij}$
(the free energy difference between the bound and unbound state).

The system is considered to be well-mixed, i.e. diffusion of all species
is assumed to occur on a fast time scale compared to reaction time scales.
Furthermore, production and degradation are assumed to be absent.
A configuration can then be characterised solely by the concentrations of
all species $c_i$ and of all dimers $c_{ij}$ at a given time. In addition,
the following conservation laws hold for every species $A_i$:
\begin{equation}
\overline{c}_i = c_i + \sum_j c_{ij}
\label{conservation}
\end{equation}
with $\overline{c}_i$ the constant total concentration of a species.

We consider mass-action kinetics, so the concentrations evolve in time
according to
\begin{equation}
\label{eq:kineq} 
\frac{{\rm d} c_i}{{\rm d} t}  = 
\sum_j \overline{K}_{ij} c_{ij} - \sum_j K_{ij} c_i c_j,
\end{equation}
and we are interested in the stationary solution ${\rm d} c_i/{\rm d}
t = 0$. A general theorem on Mass Action Reaction Networks, the so-called
{\it Deficiency Zero Theorem} \cite{fein87}, guarantees that the system
(\ref{reactions}) has a unique stationary point, irrespective of the
values of ${K}_{ij}$ and $\overline{K}_{ij}$. We omit the proof of this
statement, it amounts to calculating the deficiency of the reaction network,
which is an integer easily obtainable from the network topology (for more
details see \cite{fein87}).  Given a set of rates $\overline{K}_{ij}$ and
$K_{ij}$ and some initial concentrations, it is always possible to compute
the relaxation to equilibrium by solving the ODE in Eq.~(\ref{eq:kineq})
numerically, for instance through discretization in small time steps
$\Delta t$. However, this is a very costly procedure and thus very
unpractical for large networks.  In addition the discretization brings
some errors scaling as powers of $\Delta t$, which accumulate during the
calculation. We show here that the problem of finding the stationary
point can be reformulated as an iterative problem, which is much
more efficient and does not involve discretization approximations.

The detailed balance condition (Eq.~(\ref{DB})) provides some freedom
in choosing the forward and reverse rates. Only their ratio needs
to be fixed to guarantee convergence to the stationary point. One
particularly interesting choice is $\overline{K}_{ij}  = 1$, and
thus $K_{ij}= e^{\Delta G_{ij}/RT}$. Substituting these values in
Eq.(\ref{eq:kineq}) while setting ${\rm d} c_i/{\rm d} t = 0$ and using
Eq.~(\ref{conservation}) one finds:
\begin{equation}
\label{eq:kineq2} 
\overline{c}_i - c_i - \sum_j e^{\Delta G_{ij}/RT} c_i c_j = 0
\end{equation}
which we rewrite as
\begin{equation}
\label{eq:iter} c_i = 
\frac{\overline{c}_i}{1 + \sum_j e^{\Delta G_{ij}/RT} c_j}
\equiv T_i (c_1, \ldots c_N).
\end{equation}
where the right hand side defines a function $T$ from the N-dimensional
space of concentrations $\vec c = (c_1, \ldots c_N)$ into itself. 

Equations~(\ref{eq:iter}) are a set of $N$ non-linear equations which
must be solved to find the $c_i$.  Using vector notation we write
Eq.~(\ref{eq:iter}) as $\vec{c} = \vec{T} (\vec{c})$. A possible way to
solve this set is to use an iterative approach.  Starting from an initial
guess $\vec{c}^{\, (0)}$, one can repeatedly apply the map $\vec{T}$
to obtain $\vec{c}^{\, (1)} = \vec{T} (\vec{c}^{\, (0)})$, \ldots,
$\vec{c}^{\, (k+1)} = \vec{T} (\vec{c}^{\, (k)})$, but it remains
to be proven that the process converges to the fixed point. Indeed,
the convergence in time to a unique stable fixed point of the kinetic
equations (\ref{eq:kineq}) (according to the Zero Deficiency Theorem) does
not {\it a priori} imply the convergence of the iterated map. However,
this convergence is guaranteed by a fixed point theorem for iterated maps,
which we discuss next.

\subsection{Contraction maps}

The Banach fixed-point theorem \cite{gran03} guarantees the existence and
uniqueness of fixed points for a class of maps $\vec{c} \mapsto \vec{T}
(\vec{c})$ of a metric space into itself. A map is said to be a {\it
contraction map} if any pair of arbitrary points is mapped to a pair of
points that are closer to each other, i.e. if for any two given points
$\vec c$ and $\vec{c}{\, '}$ in some subset $\Omega$ of a metric space
for which $\vec{T} : \Omega \to \Omega$, one has:
\begin{equation}
d(\vec{T}(\vec{c}), \vec{T} (\vec{c}{\, '}))
\leq q \, d(\vec{c}, \vec{c}{\, '})
\end{equation}
with $d$ the metric (distance function) on the metric space, and with a
so-called {\it Lipschitz constant} $q < 1$.

In order to prove that a map is a contraction one has to construct a
suitable distance. Indeed, a map can be a contraction according to one
distance, but not according to another one.  We illustrate this from
a simple one-dimensional example. Let us thus consider the map of the
interval $[0,\overline{c}]$ into itself defined by
\begin{equation}
T(c) = \frac{\overline{c}}{1+K c}
\label{supp:1d}
\end{equation}
where $\overline{c} > 0$ and $K>0$. This map has a unique fixed
point since the quadratic equation $c(1+Kc)=\overline{c}$ has a single
positive solution.  In general this map is not a contraction for the
usual Euclidean distance $d_E(a,b) \equiv |a-b|$. Take for instance
$\overline{c}=2$, $K=1$ and $c=0.1$, $c'=0.2$. One has $d_E (c,c') =
0.1$, whereas $d_E (T(c),T((c')) = 0.1515\ldots$, which shows that two
points can be mapped further apart from each other.

We note that for any $c,c' \geq 0$ one has
\begin{eqnarray}
&&d_E (T(T(c)) , T(T(c')) ) = 
\left| 
\frac{\overline{c}}{1+\frac{K\overline{c}}{1+K c}} -
\frac{\overline{c}}{1+\frac{K\overline{c}}{1+K c'}} 
\right| \nonumber \\
&=& \frac{K \overline{c}|c-c'|}{(1+K(c+\overline{c}))(1+K(c'+\overline{c}))} 
\leq q |c-c'|
\label{1dcontraction}
\end{eqnarray}
where 
\begin{equation}
q = \frac{K \overline{c}}{(1+2 K \overline{c})^2} < 1
\label{q1d}
\end{equation} 
for any values of $K$ and $\overline{c}$.  (\ref{1dcontraction}) and
(\ref{q1d}) together imply that $c \mapsto f(c) \equiv T(T(c))$ is a
contraction map. Existence of a unique fixed point for contraction maps
is guaranteed by the

{\bf Banach fixed point theorem} -
{\it Let $(X,d)$ be a non-empty complete metric space. Furthermore, let
$f: X \rightarrow X$ be a contraction mapping on $X$, i.e. there exists
some $q < 1$ (called the Lipschitz constant) such that $d(f(x),f(y))
\le q \, d(x,y)$ for all $x,y \in X$. Then the map $f$ has only one
fixed point in $X$, and moreover, for any starting point $x_0 \in X$,
the sequence $\{x_n\}$ defined by $x_n = f(x_{n-1})$ converges towards
this fixed point.
}

The problem with the Euclidean distance $d_E(\vec x, \vec x') =
\sqrt{\sum_{i=1}^N (x_i-x_i')^2}$ persists in higher dimensions, as it is
possible to find $\vec{c}$, $\vec{c}^{\, \prime}$ for which $d_E(\vec{T}
(\vec{c}), \vec{T} (\vec{c}^{\, \prime})) > d_E(\vec{c}, \vec{c}^{\,
\prime})$.
Therefore, we chose a different strategy.  To use the Banach fixed point
theorem in higher dimension, we constructed an appropriate metric for
which we can explicitly prove that the map $\vec{c} \mapsto \vec{T}
(\vec{c})$ is a contraction. In the one dimensional case the metric is:

\begin{equation}
d (c,c') = \frac{|c-c'|}{c+c'}
\label{dm_1d}
\end{equation}
We prove later that this indeed defines a metric for its high-dimensional
generalization, first we will consider the one-dimensional case. We note that the presence of the
denominator in (\ref{dm_1d}) may produce a singularity when $c,c'
\to 0$. To avoid this problem we restrict ourselves to the interval
$X=[T(\overline{c}),\overline{c}]$. It can be shown easily that $T$
maps $X$ into itself and that $0 < T(\overline{c}) < \overline{c}$
so that the metric (\ref{dm_1d}) is always well-defined. $X$ is also
a compact space. Roughly speaking a space is compact if there
are no points ``missing" from it, either inside it or at the boundary. A
close interval $[a,b]$ is compact. Also a rectangle or a cube in two
and three dimensions are compact, provided all points in the borders
are included. A hypercube, the $N$-dimensional analog of a cube,
is also compact. Recall that given $N$ pairs of numbers $(a_i, b_i)$
(with $a_i < b_i$) a hypercube $X$ contains all points $(x_1, x_2,\ldots
x_N)$ such that $a_i \leq x_i \leq b_i$. We have
\begin{eqnarray}
&& d (T(c),T(c')) = \frac{K|c-c'|}{2+K(c+c')} \nonumber\\ 
&=& \frac{K(c+c')}{2+K(c+c')} \, d(c,c') \leq q \, d(c,c')
\end{eqnarray}
where
\begin{equation}
q = \frac{K \overline{c}}{1+K \overline{c}} \leq 1
\end{equation}
for any value of $K$ and $\overline{c}$. This shows that according to
the metric (\ref{dm_1d}) the map $c \mapsto T(c)$ is a contraction.
In the following we will introduce a metric which is a higher dimensional
generalization of (\ref{dm_1d}).

\subsection{Higher dimensional map}

Consider any $N \times N$ matrix with non-negative entries
($K_{ij} \geq 0$). We prove that the map defined in Eq.~(\ref{eq:iter}) is
a contraction in the space $X$ of $N$-dimensional vectors $\vec c = (c_1,
c_2 \ldots c_N)$ such that all elements are $\varepsilon_i \leq c_i \leq
\overline{c_i}$, where $\overline{c_i}$ are fixed. Here we have defined
\begin{equation}
\varepsilon_i \equiv \frac{\overline{c_i}}{1+\sum_j K_{ij} \overline{c_j}} > 0
\end{equation}
It is easy to show that $\vec{T}$ maps $X$ into itself. $X$ is also
compact.

We consider the metric defined as
\begin{equation}
d(\vec{c}, \vec{c}{\, '}) = \max_{1 \leq i \leq N} \frac{|c_i - c_i'|}{c_i + c_i'}
\label{eq:metr2} 
\end{equation}
which is a higher dimensional generalization of (\ref{dm_1d}).  To prove
that $\vec{T}$ is a contraction we have to show that for any two
points in $X$, say $\vec{c}$ and $\vec{c}{\ '}$, one has
\begin{equation}
d(\vec{T}(\vec{c}), \vec{T} (\vec{c}{\, '}))
\leq q \, d(\vec{c}, \vec{c}{\, '})
\label{contraction}
\end{equation}
with Lipschitz constant $q < 1$. We first show
that (\ref{eq:metr2}) has the mathematical properties of a distance and
then that (\ref{contraction}) holds.

\subsubsection{Eq.~(\ref{eq:metr2}) defines a metric}

To show that $d()$ as defined by Eq.~(\ref{eq:metr2}) is a metric on
the metric space $X$, we need to prove that for every $\vec{c}$,
$\vec{c}{\, '}$ and $\vec{c}{\, ''}$ in $X$
\begin{itemize}
\item[a)] $d(\vec{c}, \vec{c}{\, '})  \geq 0$
and  $d(\vec{c}, \vec{c}{\, '}) = 0$ {\rm iff} $\vec c = \vec{c}{\, '}$
\item[b)] $d(\vec{c}, \vec{c}{\, '})  = d(\vec{c}{\, '}, \vec{c}) $
\item[c)] Triangle inequality: $d(\vec{c}, \vec{c}{\, '}) \leq d(\vec{c}, \vec{c}{\, ''}) 
+ d(\vec{c}{\, ''}, \vec{c}{\, '})$
\end{itemize}

{\bf Proof}: a) and b) are trivial. The triangle inequality requires some
more work.  We need to prove that
\begin{equation}
\max_i \frac{|c_i - c'_i|}{c_i+c'_i} \le
\max_i \frac{|c_i - c''_i|}{c_i+c''_i}
+ \max_i \frac{|c''_i - c'_i|}{c''_i+c'_i},
\end{equation}
We shall prove that the inequality holds for every $i$,
thus that for any non-negative $a$, $b$ and $c$ one has
\begin{equation}
\label{eq:tri}
\frac{|a-b|}{a+b} \le \frac{|a-c|}{a+c} + \frac{|c-b|}{c+b}.
\end{equation}

First of all we note that the inequality (\ref{eq:tri}) is satisfied
when $a=0$ or $b=0$ or $c=0$. It is also satisfied when two elements are
equal $a=b$, $a=c$ or $b=c$.
We have to consider then these different cases:
(1) $0 < c < b < a$,
(2) $0 < c < a < b$,
(3) $0 < b < c < a$,
(4) $0 < b < a < c$,
(5) $0 < a < c < b$ and
(6) $0 < a < b < c$.
However, the inequality (\ref{eq:tri}) is symmetric in the exchange of
$a$ with $b$. We have to prove it only for the cases  for which $a>b$:
(1) $0 < c < b < a$, (3) $0 < b < c < a$ and (4) $0 < b < a < c$.

\bigskip
(1) $0 < c < b < a$.
\medskip

The inequality (\ref{eq:tri}) becomes
\begin{equation}
\frac{a-b}{a+b} \leq \frac{a-c}{a+c} + \frac{b-c}{c+b}.
\end{equation}
which, after some elementary algebra, can be rewritten as
\begin{equation}
(b-c) [2a(b+c) + (a+c)(a+b)] \geq 0
\end{equation}
This inequality is verified in the case (1) since $b>c>0$ and $a > 0$.

\bigskip
(3) $0 < b < c < a$
\medskip

The inequality (\ref{eq:tri}) becomes
\begin{equation}
\frac{a-b}{a+b} \leq \frac{a-c}{a+c} + \frac{c-b}{c+b}
\end{equation}
and after some simple algebra we get
\begin{equation}
(c-b)(a-c)(a-b) \geq 0
\end{equation}
which is satisfied for $a > c > b > 0$.

\bigskip
(4) $0 < b < a < c$
\medskip

The inequality (\ref{eq:tri}) becomes
\begin{equation}
\frac{a-b}{a+b} \leq \frac{c-a}{a+c} + \frac{c-b}{c+b}
\end{equation}
which can be rewritten as
\begin{equation}
(c-a) [2b(a+c) + (a+b)(c+b)] \geq 0
\end{equation}
which is again satisfied for $c>a$.

This proves that the triangle inequality is satisfied. Hence $d$ defines
a metric on $X$.

\subsubsection{The inequality (\ref{contraction}) is verified}

Combining Eqs.(\ref{eq:iter}) and (\ref{eq:metr2}) we find
\begin{eqnarray}
d(\vec{T}(\vec{c}), \vec{T} (\vec{c}{\, '})) &=& \max_i
\frac{ \left| \frac{\overline{c_i}}{1+\sum_j K_{ij} c_j} - \frac{\overline{c_i}}{1+\sum_j
K_{ij} c_j'} \right| }{\frac{\overline{c_i}}{1+\sum_j K_{ij} c_j} +
\frac{\overline{c_i}}{1+\sum_j K_{ij} c_j'}}
\nonumber \\
&=& \max_i \frac{ \left| \sum_j K_{ij} (c_j-c_j') \right| }{
2+\sum_j K_{ij} (c_j+c_j') } 
\nonumber \\
&\leq&
\max_i \frac{ \sum_j K_{ij} \left| c_j-c_j' \right| }{
2+\sum_j K_{ij} (c_j+c_j') }
\label{contr1}
\end{eqnarray}

To proceed further we make use of the inequality:
\begin{equation}
\frac{\sum_l a_l}{\sum_l b_l} \leq \max_l \left( \frac{a_l}{b_l}\right)
\label{ineq1}
\end{equation}
valid for $a_l, b_l > 0$. We prove this inequality in the case
\begin{equation}
\label{eq:sumineq}
\frac{a_1+a_2}{b_1+b_2} \le \max \left(\frac{a_1}{b_1}, \frac{a_2}{b_2} \right)
\end{equation}
from which (\ref{ineq1}) follows easily by repeatedly applying
(\ref{eq:sumineq}).  To verify (\ref{eq:sumineq}) consider ${a_1}/{b_1}
\geq {a_2}/{b_2}$. From this we have ${a_1}{b_2} \geq {a_2}{b_1}$ and
${a_1}{b_2} + {a_1}{b_1} \geq {a_2}{b_1} + {a_1}{b_1}$, which implies
${a_1}/{b_1} \geq (a_1+a_2)/(b_1+b_2)$ and proves (\ref{eq:sumineq}).

Note that from (\ref{ineq1}) it is immediately clear that
\begin{eqnarray}
&&\max_i \frac{ \sum_j K_{ij} \left| c_j-c_j' \right| }{
2+\sum_j K_{ij} (c_j+c_j') }
\leq \max_i \frac{ \sum_j K_{ij} \left| c_j-c_j' \right| }{
\sum_j K_{ij} (c_j+c_j') } \nonumber\\
&=& 
\max_i \max_j \frac{K_{ij} \left| c_j-c_j' \right| }{K_{ij}(c_j+c_j')}
= d(\vec{c} , \vec{c}{\, '})
\end{eqnarray}
which proves that
\begin{equation}
d(\vec{T}(\vec{c}), \vec{T} (\vec{c}{\, '})) \leq
d(\vec{c} , \vec{c}{\, '})
\end{equation}
which is close to the desired result, but it does not suffice, because our
aim is to prove that there exists a $q$ that is {\it strictly smaller}
than $1$ for which (\ref{contraction}) is satisfied.

To do that we proceed as follows.  Let us define $k_{\max} = \max_{i,j}
K_{ij}$ and $\overline{c}_{\max} = \max_i \overline{c_i}$.  We rearrange
the denominator of the last term of (\ref{contr1}) by adding and
subtracting the same term as follows
\begin{eqnarray}
&& 
2+\sum_j K_{ij} (c_j+c_j') = 
\sum_j \frac{2}{N} \left(1-\frac{K_{ij}}{k_{\max}} 
\frac{c_j+c_j'}{2 \overline{c}_{\max}}\right)
\nonumber \\
&+&
\sum_j K_{ij}\left(1+\frac{1}{N k_{\max} \overline{c}_{\max}}\right) (c_j+c_j') 
\nonumber \\
&\geq&
\left(1+\frac{1}{N k_{\max} \overline{c}_{\max}}\right) 
\sum_j K_{ij} (c_j+c_j') 
\nonumber \\
&=& \frac 1 q
\sum_j K_{ij} (c_j+c_j')
\label{contr2}
\end{eqnarray}
where we have defined
\begin{equation}
q \equiv \left( 1+\frac{1}{N k_{\max} \overline{c}_{\max}} \right)^{-1} < 1
\label{lipsc}
\end{equation}
In deriving (\ref{contr2}) we have used $K_{ij} \leq k_{\max}$ and $c_j
+c_j' \leq 2 \overline{c}_{\max}$ which guarantees that
\begin{equation}
1-\frac{K_{ij}}{k_{\max}} \frac{c_j+c_j'}{2 \overline{c}_{\max}} \geq 0
\end{equation}

Finally, combining (\ref{contr2}) and (\ref{contr1}), and followed by the inequality (\ref{ineq1}), we obtain
\begin{eqnarray}
&&d(\vec{T}(\vec{c}), \vec{T} (\vec{c}{\, '})) 
\leq 
q \max_i \frac{\sum_j K_{ij} \left| c_j-c_j' \right|}{\sum_j K_{ij} (c_j+c_j')}
\nonumber \\
&\leq& q \max_i \max_j \frac{K_{ij} \left| c_j-c_j' \right| }{K_{ij}(c_j+c_j')}
\leq q d(\vec{c}, \vec{c}{\, '}),
\end{eqnarray}
concluding our proof to confirm that (\ref{contraction}) indeed holds
for all $c,c'$ in $X$. A Lipschitz constant $q$, which is required by the
Banach fixed point theorem, is then given by (\ref{lipsc}).

\section{Hybridization reactions in human transcriptome}
\label{sec:mRNA}

Having proven the convergence of the iterative algorithm for generic
hetero-dimerization networks, irrespective of the values of the rates,
we proceed with a specific example from biology.  In this example the
chemical species are messenger RNA (mRNA) fragments taken from the
human genome databank (details below). 

\begin{figure}[!ht]
\begin{center}
\includegraphics[width=7cm]{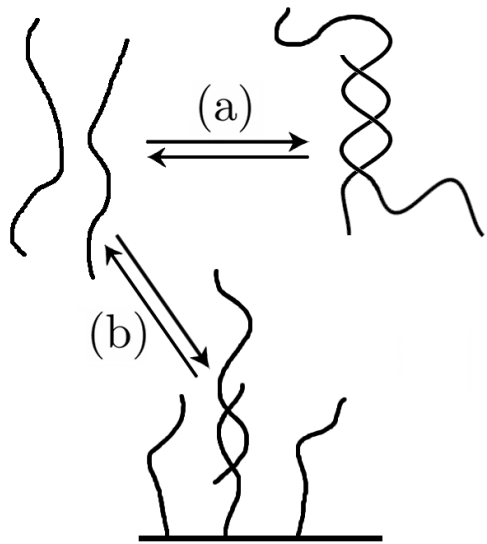}
\end{center}
\caption{(a) Hybridization reaction in solution between partially
complementary mRNA strands. (b) Hybridization reaction between mRNA
strand from solution and substrate-based microarray DNA probe.}
\label{hybrid_dimer}
\end{figure}

To clarify the importance of this example we recall briefly some facts.
In order to understand the function of the genes in an organism, it is
essential to know under which conditions (or in which cell types in
a multicellular organism) they are expressed, i.e. transcribed into
single stranded mRNA. High throughput devices such as DNA microarrays 
\cite{bald02} have
been extensively used for this type of analysis because they provide
information on the whole transcriptome, the set of all RNAs produced
by the cells by transcription, on a single experiment. On a microarray
the complement of one or more fragments of a specific transcript,
referred to as probes, are used as reporters. The probe sequences are
covalently linked on a solid surface in spots. A solution containing
the mRNA extracted from cells is deposited on the microarray surface. A
transcript in solution with a sequence complementary to that of the
probe tends to bind to it, a process known as hybridization, which is
illustrated in Fig.~\ref{hybrid_dimer}(b). 

Typically, a large number of genes is transcribed simultaneously
in cells, hence mRNA extracted from biological samples contains
many different sequences that cover a broad range of concentrations,
reflecting the broad differences in expression levels. Single stranded
nucleic acids in solution tend to bind to sequences which are partially
complementary to them, resulting in a double helical fragment, as illustrated
in Fig.~\ref{hybrid_dimer}(a). The hybridization between partially
complementary fragments in solution ``competes" with the hybridization
to the sequences at the microarray surface.

Consider a single stranded mRNA fragment $t$ transcribed from a given
gene. If the solution contains another transcript $t'$ which has a strong
tendency to hybridize to the first fragment, both or only one of the
two sequences may get significantly ``depleted" from the solution.
If the binding is strong enough, duplex formation may continue until
almost all fragments of the least abundant type are hybridized. 

As pointed out by several papers \cite{carl06,bind06,burd06,halp06} the
presence of hybridization in solution may lead to an underestimation
of expression levels from microarray data analysis. It is therefore
important to be able to quantify its effect.  This is the aim of
this example discussed here. The equilibrium and kinetics of mutual
hybridization between DNAs was studied before~\cite{horn06}, but only
for systems with about $N\sim 10^2$ sequences.

\subsection{The sample}

For the computation we considered a database containing
$33,457$ human mRNA's sequences downloaded from
\url{ftp.ncbi.nih.gov/refseq/H_sapiens/mRNA_Prot/}, file {\tt
human.rna.fna}.  These transcripts have an average length of several
thousand nucleotides. However, in typical biochemical assays, as for
instance in microarray experiments \cite{bald02}, the transcripts
are present in shorter fragments of various lengths.  We have chosen
to divide the sequences into fragments of $48$ nucleotides, starting
from the $5'$ end of the transcript. The first fragment starts thus
from nucleotide $n_1=1$ of the given transcript.  The $m$-th fragment
starts at nucleotide position $n_m=n_{m-1}+8$, i.e. with a shift of $8$
nucleotides from the previous one. This procedure avoids artifacts due
to the exact location of the fragmentation point. For a transcript of
length $L$ the fragmentation produces thus $L/8$ fragments, rounded down.

For the transcripts analyzed, the fragmentation produces in total
$N=3,150,659$ different $48$-mers. To each of these an initial
concentration $\overline{c}_i$ is assigned, such that fragments
originating from the same transcript are given the same concentration.
Input concentrations were obtained from DNA microarray data of Human
mRNA in different tissues, using the outputs from the data analysis
algorithm discussed in Ref.~\cite{muld09}.  Typical concentrations range
from a few picomolars (pM) for low expressed genes, to nanomolar (nM)
for the highly expressed genes.

The hybridization free energies $\Delta G_{ij}$, used in
Eqs.~(\ref{eq:iter}) were computed using the nearest-neighbor model
\cite{bloo00}, which assumes that the stability of the double helix
depends on the identity and orientation of neighboring base pairs. The
total free energy of a hybridizing strand is obtained as the sum of
$10$ independent parameters accounting for hydrogen bonding and stacking
interactions. In our computations we used the RNA/RNA parameters given
in \cite{xia98}, at $1$M [Na+] and a temperature $T=55 \celsius$.

\begin{figure}[!ht]
\begin{center}
\includegraphics[width=8cm]{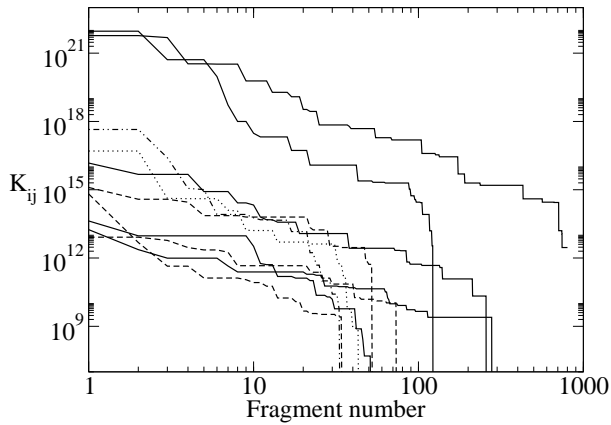}
\end{center}
\caption{Plot of $K_{ij}$ for $10$ randomly selected
fragments. The data are shown in decreasing order.  It is seen in this
figure that for each fragment, there are typically a few other fragments
in the solution with which there is a significant interaction. Lists were
generated containing all fragments which were complementary for at least
8 consecutive nucleotides.  The number of `partners' within a given free
energy range is growing roughly exponentially, similar to what would be
expected for randomly generated sequences. However, the Boltzmann factors
$K_{ij} = e^{\beta \Delta G_{ij}}$ decrease faster, such that the highest
few factors still dominate the sum $\sum_j c_j K_{ij}$ in the iterative
scheme. Consequently, we can approximate this sum by truncating the list.}
\label{Fig_freed}
\end{figure}

\subsection{Efficient construction of interaction matrix}

The iterative scheme presented above has a computational cost of order
$N^2$: for each of the $N$ equations of (\ref{eq:iter}) one has to compute
the sum of $N$ terms. However, the analysis of the terms entering in
the sum in the denominator of Eq.~(\ref{eq:iter}) shows that this sum
is dominated by a few terms corresponding to the highest values of the
hybridization free energy. Figure~\ref{Fig_freed} shows a plot of the
$K_{ij}$ for some randomly selected species $i$, as a function of $j$
ranked in decreasing order. The Boltzmann factors $K_{ij} = e^{\beta
\Delta G_{ij}}$ decay rapidly as a function of $j$.  As an approximation
we kept only the first ten dominant terms for each $K_{ij}$, estimating
that the typical error on the results is of a few percent.  This improves
the memory requirements of any implementation, and hence allows a much
greater number of sequences to be present in our calculations.

To build up the matrix elements efficiently we generate a list of all
possible sequences of length $l=8$ (this list has size $4^l$ and we refer
to it as the primary list). We then run through all the mRNA sequences
and generate an index vector which maps each position on a corresponding
element of the primary list.  This is an operation of order $N$. Having
found two sequences $i$ and $j$ that are complementary for a stretch of
length $l=8$, we can check if this complementarity can be extended to a
longer stretch. This method is still of computational complexity of order
$N^2$, however, with a small prefactor compared to a full complementarity
matching. With the used method one ignores complementarity for stretches
shorter than $l=8$ nucleotides, but these sequences would not be expected to
cause significant hybridization anyways. The implementation can become very
efficient by using binary operations: the four nucleotide types are encoded
into two bits, complementarity can then be easily checked by a bitwise
XOR operation.

\begin{figure}[t]
\begin{center}
\includegraphics[width=8cm]{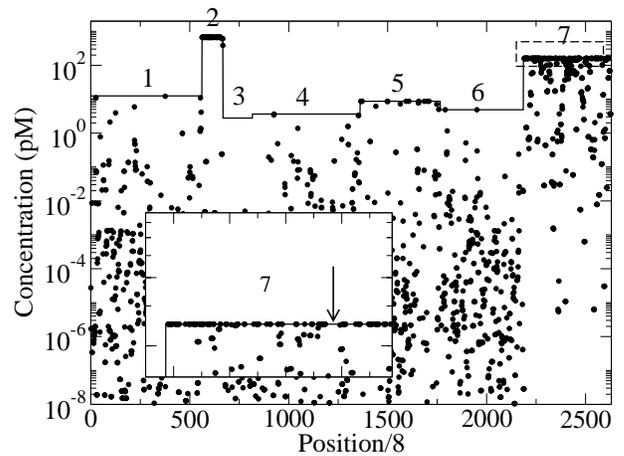}
\end{center}
\caption{Initial concentrations ($\overline{c}_i$, solid
line) and equilibrium single stranded concentration ($c^*_i$, circles)
for seven selected transcripts. Fragments with a high total concentration
are typically unaffected by hybridization in solution, but some of the
fragments with a low total concentration get significantly depleted.
Inset: zoom of the dashed zone around transcript $7$. The arrow shows
a region where significant depletion has taken place.}
\label{Fig_conc_transcripts}
\end{figure}

\subsection{Results}

Once the initial concentrations $\overline{c}_i$ of the $N$ fragments
are fixed, we repeatedly apply the map defined in Eq.~(\ref{eq:iter}). As
proven before, the iterative procedure converges to a unique fixed point,
representing the equilibrium concentration of fragments which are not
hybridized.

In practice, the convergence criterion has been chosen such that the
distance in concentration (using the distance $d(\vec{c},\vec{c}') =
\sum_i |c_i-c_i'|/|c_i+c_i'|$ for convenience) between two successive
iterations is smaller than a given small value: $d(\vec{c}^{\, (k)},
\vec{c}^{\, (k+1)}) \leq \varepsilon$ with $\varepsilon=1$ Typically about
$10^2$ iterations are sufficient to guarantee this level of accuracy.

Figure~\ref{Fig_conc_transcripts} shows a typical output of the
computation for $7$ randomly chosen transcripts of varying lengths. The
fragments are ordered as they are generated from the fragmentation
procedure described above, thus the horizontal scale should be multiplied
by a factor $8$ to have the length in nucleotides.  The thin solid
line corresponds to the total concentration $\overline{c}_i$ which,
as mentioned before, is chosen to be constant for each transcript.
For the transcripts shown in Fig.~\ref{Fig_conc_transcripts} the initial
concentrations range from $2.8$ pM (for transcript $3$) to $670$ pM
(for transcript $2$).  The points denote the equilibrium concentration
of single strands $c^*_i$, computed from the iterative algorithm. The
figure shows different types of behaviors for the transcripts. Transcript
$2$, which had the highest initial concentration, is weakly affected
by hybridization with other fragments and most of the fragments have
a free concentration very close to the initial one ${c}_i \approx
\overline{c}_i$. Transcript $7$ is moderately affected by hybridization
with other targets. A few of its fragments strongly hybridized with
complementary partners in solution such that the concentration of free
strands can drop of several orders of magnitude.  The other transcripts,
whose initial concentration was of few picomolars, are strongly affected
and for almost all fragments, the free concentration is much lower than
the initial concentration, ${c}_i \ll \overline{c}_i$.

As mRNA has typically a length of several thousands of nucleotides,
there are many possible ways in which probes can be selected from it
(most microarrays use as probes of 20-50 nucleotides). Probe design
is a fundamental step in the realization of a DNA microarray. As an
example, the results of Fig.~\ref{Fig_conc_transcripts}(inset) suggest
that a probe selected in the region marked by the arrow on transcript
$7$ can significantly underestimate the true expression level of the
transcript. As the transcript fragment strongly hybridizes in solution
(reaction (a) of Fig.~\ref{hybrid_dimer}) a small concentration of single
strand is available for hybridizing to the microarray surface (reaction
(b) of Fig.~\ref{hybrid_dimer}).  Based on these results, an additional
important criterion for probe design should be the avoidance of strongly
hybridizing transcriptomic regions.

Computations have been performed on different samples, differing
by the values of the $\overline{c}_i's$, reflecting the different
expression levels expected in different human tissues. The details of
the computations will be presented elsewhere (F. Berger, M. G. A. van
Dorp and E.~Carlon, unpublished). As in the example above, we find in
the transcriptome some regions strongly affected by mutual hybridization.

\section{Convergence of the iterative procedure}
\label{sec:convergence}

\begin{figure}[t]
\begin{center}
\includegraphics[width=9cm]{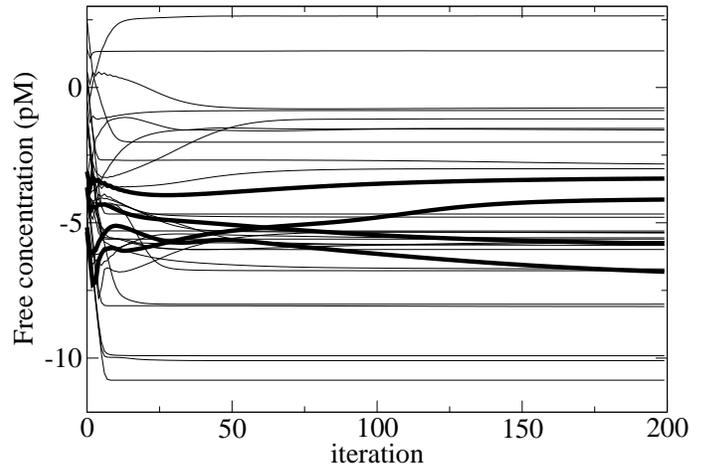}
\end{center}
\caption{
Convergence of $32$ species in the human transcriptome analysis. The
horizontal axis is for the number of iterations, while the vertical
axis shows the concentration for each of the $32$ probes at each
iteration. Concentrations appear to be fairly stable at $200$ iterations,
suggesting a fair degree of convergence. The four thicker lines correspond
to four concentrations that display slow convergence.
} 
\label{Fig_convtest}
\end{figure}

In ths section we discuss the speed of convergence of the iterative
algorithm for the specific example of mutual hybridizations in the human 
transcriptome and more in general for a generic hetero-dimerization network.

\subsection{Convergence for human transcriptome hybridization analysis}

Figure~\ref{Fig_convtest} shows the concentrations $c_i^{(k)}$ of $32$
randomly selected species as a function of the iteration number $k$
for the reaction network of hybridizing mRNA fragments discussed in
Section~\ref{sec:mRNA}. One notices that the convergence to the stationary
value is attained for the majority of species after about $50$ iterations.
However the speed of convergence varies for the different species, and
in particular for three species in Fig.~\ref{Fig_convtest} convergence
is much slower than average (thick lines). In all cases shown, after
about $k \approx 200$ iterations $c_i^{(k)}$ has become mostly stationary.

The Banach theorem provides an estimate of the convergence rate given
by $q$, as defined in Eq.~(\ref{lipsc}). However this is not a very
practical bound for convergence, as even in cases where just a few species bind
very strongly to each other, we easily have $k_{\max} \overline{c}_{\max}
\gg 1$. More realistic convergence rates can be obtained from the linear
stability analysis around the fixed point.

\subsection{Linear stability analysis predictions for convergence}

Given $N$ dimerizing species, we construct the matrix
\begin{equation}
\tilde{J}_{mn} \equiv  - \frac{\partial T_m (\vec{c}^{\, *})}{\partial c_n} =
\frac{K_{mn} \overline{c}_m}{(1+\sum_j K_{mj} c^*_j)^2}
= \frac{K_{mn} {c}^*_m}{1+\sum_j K_{mj} c^*_j}
\label{jacobian}
\end{equation}
where $\vec{T} (\vec c )$ is the iterated map defined by
Eq.~(\ref{eq:iter}) and $\vec{c}^{\, *}$ denotes the fixed point.
Note that $\tilde{J}$ is obtained from the Jacobian matrix, by swapping
the signs of all entries.  The matrix $\tilde{J}$ is non-negative
in the sense that for all its elements $\tilde{J}_{mn} \geq 0$. In
the following we will use the notation $\vec{v} < \vec{w}$ if the
inequality holds for all elements of the two vectors $\vec{v}$ and
$\vec{w}$.  For a non-negative matrix $J$, a common extension to the
Perron-Frobenius theorem \cite{gant59} guarantees that there exists a
(not necessarily unique) largest eigenvalue $r>0$ whose eigenvector
$\vec\phi$ is non-negative, $\vec\phi \geq 0$. This eigenvector is known
as the Perron-Frobenius vector.  The largest eigenvalue $r$ determines
the slowest convergence rate of the iterative scheme.

We have proven that the map $T$ is a contraction, hence necessarily $r<1$.
We can however derive a stronger bound as follows. From $(\ref{jacobian})$
one derives
\begin{equation}
\left( \tilde{J} \vec{c}^{\, *}\right)_m =  \sum_n \tilde{J}_{mn} {c}^*_n =
\frac{\sum_n K_{mn} {c}^*_n}{1+\sum_j K_{mj} c^*_j} c_m^* 
\equiv \alpha_m c_m^*
\label{jacob_ineq}
\end{equation}

Consider now $\tilde{J}^T$, the the transpose of $\tilde{J}$.
The transpose has the same eigenvalues as $\tilde{J}$, but a different
eigenvector. The Perron-Frobenius theorem applies also to $\tilde{J}^T$
for which $\tilde{J}^T
\vec{\phi}{\, '}
= r \vec{\phi}{\, '}$ and $\vec{\phi}{\, '} \geq 0$. One has
\begin{equation}
r \vec{\phi}{\, '} \cdot \vec{c}^{\, *} = \tilde{J}^T \vec{\phi}{\, '} \cdot
\vec{c}^{\, *}
=  \vec{\phi}{\, '} \cdot \tilde{J} \vec{c}^{\, *} 
\label{final_ineq}
\end{equation}
where the dot indicates the scalar product. 

Working out this scalar product and using Eq.~(\ref{jacob_ineq}) one finds
\begin{equation}
r \sum_m \phi'_m {c^*_m} = \sum_m \phi'_m \alpha_m {c^*_m}
\leq \max_n \{ \alpha_n \} \sum_m \phi'_m {c^*_m}
\label{eq:rbound}
\end{equation}
where we have used the fact that $\phi'_m\geq 0$ and $c^*_m \geq 0$ 
Equation~(\ref{eq:rbound}) shows that
\begin{equation}
r \leq \max_m \frac{\sum_j K_{mj} {c}^*_j}{1+\sum_j K_{mj} c^*_j} < 1.
\label{qlin}
\end{equation}
For most networks, especially those where most values $K_{mn}$ are
small and only a few ones are very large, this is a better bound to
the convergence rate compared to that guaranteed by Banach's theorem
(Eq.~(\ref{lipsc})) as:
\begin{equation}
\max_m \frac{\sum_j K_{mj} {c}^*_j}{1+\sum_j K_{mj} c^*_j} = 
\frac{1}{1+\frac{1}{\max_m \sum_j K_{mj} {c}^*_j}} < q
\label{bound_rate}
\end{equation}
where $q$ is the Lipschitz constant given by Eq.~(\ref{lipsc}).

The largest eigenvalue of the Jacobian matrix provides the slowest rate
of convergence.  The example of Fig.~\ref{Fig_convtest} shows that the
convergence rate may differ for different species. We provide here some
simple insights on possible origins of the slow convergence.

Consider first reaction networks for which the dimerization process is
very weak, so that the equilibrium concentrations $c_m^*$ differ only
weakly from the total concentration $\overline{c}_m$. In this case $K_{mn}
c^*_n \ll 1$, hence Eq.~(\ref{qlin}) implies fast convergence.

More interesting is the case of strongly interacting networks, where
slow relaxation to equilibrium is expected. To illustrate this, consider
two strongly interacting species for which dimerization is so strong
that interaction with the rest of the network can be neglected in first
approximation. This two species system is described by the equations
\begin{eqnarray}
c_1 = \frac{\overline{c}_1}{1+K c_2} \\
c_2 = \frac{\overline{c}_2}{1+K c_1}
\end{eqnarray}
where for simplicity we discard the possibility of self-dimerization
and $K=K_{12}=K_{21}$. The diagonal elements of the Jacobian vanish
($\tilde{J}_{11}=\tilde{J}_{22}=0$), therefore the rate of convergence
(eigenvalues of $\tilde{J}$) is given by
\begin{equation}
\lambda = \pm \sqrt{\tilde{J}_{12}\tilde{J}_{21}}
\end{equation}

We consider now two different cases: (1) $\overline{c}_1 = \overline{c}_2
= \overline{c}$ and $K \overline{c} \gg 1$ and (2) $\overline{c}_1 \gg
\overline{c}_2$ and $K \overline{c}_2 \gg 1$.

The case (1) corresponds to the two species having the same initial concentration
and strongly dimerizing to each other. In equilibrium $c^*_1 = c^*_2 \ll \overline{c}$.
Some simple algebra shows that the eigenvalues of the Jacobian become
\begin{eqnarray}
\lambda = \pm 1 + {\cal O} \left( \frac{1}{\sqrt{K \overline{c}}} \right)
\end{eqnarray}
which implies a very slow convergence, as $|\lambda|$ is close to $1$.
In the case (2) one finds
\begin{eqnarray}
\lambda = \pm \sqrt{\frac{\overline{c}_2}{\overline{c}_1}} 
\end{eqnarray}
which is small when $\overline{c}_1 \gg \overline{c}_2$. This implies very fast
convergence.

If this type of problems would be encountered in large networks, we suggest to
solve them by computing analytically the equilibrium values for the subnetwork
containing only these two species. Putting the resulting equilibrium
concentrations as initial concentrations in the full network, the iterative
scheme should converge fast. Finally, we remark that problematic
cases can easily be detected by checking whether $\overline{c}_1 \approx
\overline{c}_2$. If this is not the case, convergence will necessarily be fast. 

\section{Conclusions}
\label{sec:conclusion}

We have described a novel algorithm to efficiently compute the equilibrium
concentrations for a class of chemical reaction networks. The core
part of our algorithm is an iterative procedure, which we have shown
to converge to the unique stable point of the system. Furthermore, we
have analysed the convergence properties to conclude that convergence
is typically fast, except in certain problematic cases. These problems
turn out to be easy to detect, and the solution to these problems by
analytically solving subnetworks could eventually be implemented.

The inspiration for considering this problem can be traced back to the analysis
of physical models describing experiments with DNA microarrays. Consequently, we
have implemented our algorithm specifically to test whether it can be used to
find the equilibrium concentrations for a system of this size, where of the
order of $10^6$ different RNA fragments form a large reaction network. We found
that convergence was quick, on the order of at most a few hundred iterations,
and that the full computation took only a few minutes on a mainstream desktop PC.

In the present work we have restricted our analysis to hetero-dimerization networks
with mass action kinetics. It would be very interesting to expand these iterative
algorithms to a wider class of chemical reaction networks, particularly for the
case of genetic regulatory networks \cite{dejo02}. One factor limiting the study
of these networks is that often the values of reaction rates are not known apart
for very few well studied cases~\cite{dejo02}. The steady state analysis has to
be repeated for various input rates, therefore it is very important to have fast
algorithms, which could perform this analysis efficiently.

\section*{Acknowledgments}

M.v.D and E.C. are grateful to the Kavli Institute for Theoretical Physics
China (Beijing), where part of this work was done, for kind hospitality.
Discussions with G.T. Barkema and M. Fannes are gratefully acknowledged.
We acknowledge financial support from KULeuven grant OT/STRT1/09/042.


%
\end{document}